\documentclass{elsart}
\journal{New Astronomy}

\usepackage{natbib}

\usepackage{graphicx}
\usepackage{epsfig}

\usepackage{amssymb}

\begin{document}

\begin{frontmatter}



 \title{The Cosmic Infrared Background Experiment}


\author{James Bock$^{1,2}$, John Battle$^1$, Asantha Cooray$^3$, Mitsunobu Kawada$^4$}
\author{Brian Keating$^5$, Andrew Lange$^2$, Dae-Hea Lee$^6$, Toshio Matsumoto$^7$}
\author{Shuji Matsuura$^7$, Soojong Pak$^6$, Tom Renbarger$^5$, Ian Sullivan$^2$}
\author{Kohji Tsumura$^7$, Takehiko Wada$^7$, Toyoki Watabe$^4$}
\address{$^1$Jet Propulsion Laboratory, Pasadena CA 91109\\
$^2$California Institute of Technology, Pasadena CA 91125\\
$^3$Physics and Astronomy, UC Irvine, Irvine CA 92697\\
$^4$Nagoya University, Furo-cho, Chikusa-ku, Nagoya 464-8602, Japan\\
$^5$Department of Physics, UC San Diego, San Diego CA 92093\\
$^6$Korea Astronomy and Space Science Institute, Daejeon 305-348, Korea\\
$^7$ISAS, JAXA, Sagamihara, Kanagawa 229-8510, Japan\\
E-mail: James.Bock@jpl.nasa.gov}

\begin{abstract}
The extragalactic background, based on absolute measurements reported by DIRBE
and IRTS at 1.2 and 2.2 $\mu m$, exceeds the brightness derived from galaxy counts
by up to a factor 5. Furthermore, both DIRBE and the IRTS report fluctuations in the
near-infrared sky brightness that appear to have an extra-galactic origin, but are
larger than expected from local (z = 1-3) galaxies. These observations have led to
speculation that a new class of high-mass stars or mini-quasars may dominate primordial
star formation at high-redshift $(z \sim 10-20)$, which, in order to explain the
excess in the near-infrared background, must be highly luminous but produce a limited
amount of metals and X-ray photons. Regardless of the nature of the sources, if a
significant component of the near-infrared background comes from first-light galaxies,
theoretical models generically predict a prominent near-infrared spectral feature
from the redshifted Lyman cutoff, and a distinctive fluctuation power spectrum.

We are developing a rocket-borne instrument (the Cosmic Infrared Background ExpeRiment,
or CIBER) to search for signatures of primordial galaxy formation in the cosmic
near-infrared extra-galactic background.  CIBER consists of a wide-field two-color
camera, a low-resolution absolute spectrometer, and a high-resolution narrow-band
imaging spectrometer.

The cameras will search for spatial fluctuations in the background on angular
scales from $7''$ to $2^\circ$, where a first-light galaxy signature is expected
to peak, over a range of angular scales poorly covered by previous experiments.
CIBER will determine if the fluctuations reported by the IRTS arise from
first-light galaxies or have a local origin.  In a short rocket
flight CIBER has sensitivity to probe fluctuations $100 \times$ fainter than
IRTS/DIRBE, with sufficient resolution to remove local-galaxy correlations.  By
jointly observing regions of the sky studied by Spitzer and ASTRO-F, CIBER will
build a multi-color view of the near-infrared background, accurately assessing
the contribution of local (z = 1-3) galaxies to the observed background fluctuations,
allowing a deep and comprehensive survey for first-light galaxy background fluctuations.

The low-resolution spectrometer will search for a redshifted Lyman cutoff feature
between 0.8 - 2.0 $\mu m$. The high-resolution spectrometer will trace zodiacal light
using the intensity of scattered Fraunhofer lines, providing an independent measurement
of the zodiacal emission and a new check of DIRBE zodiacal dust models. The combination
will systematically search for the infrared excess background light reported in
near-infrared DIRBE/IRTS data, compared with the small excess reported at optical
wavelengths.

\end{abstract}


\end{frontmatter}

\section{The Infrared Background}
Is the infrared background (IRB) entirely produced by discrete local galaxies, or is
there an unaccounted component?  Integrated individual galaxy counts appear to fall
short of the IRB measured with absolute photometry at near-infrared wavelengths.  The
cause is either unaccounted systematic errors, or a tantalizing clue to new
cosmology: a diffuse IRB component from the early universe.  A new
population of energetic sources at high redshift \cite{Wyithe03}, \cite{Cooray04b},
\cite{Madau05} would be needed to explain the intensity of the near-infrared
background as reported by several authors.  Because the infrared background may
have profound implications for cosmology, these measurements must be confirmed by
independent means.

The IRB may be studied with absolute photometry or integrated number counts of galaxies.
Both methods are confronted with systematic errors.  With absolute photometry, one must
account for instrument emission and baselines, and remove local foregrounds, namely stars,
interstellar dust, interplanetary dust emission, at mid-infrared and far-infrared wavelengths,
and sunlight scattered by interplanetary dust ('zodiacal light'), at optical and near-infrared
wavelengths \cite{Hauser01}.  Galaxy counts could possibly under-count flux from diffuse
components or miss objects with low surface brightness, especially in ground-based infrared
surveys using spatial chopping (see \cite{Totani01} for a quantitative assessment).

\begin{figure}[h]
\centerline{\psfig{file=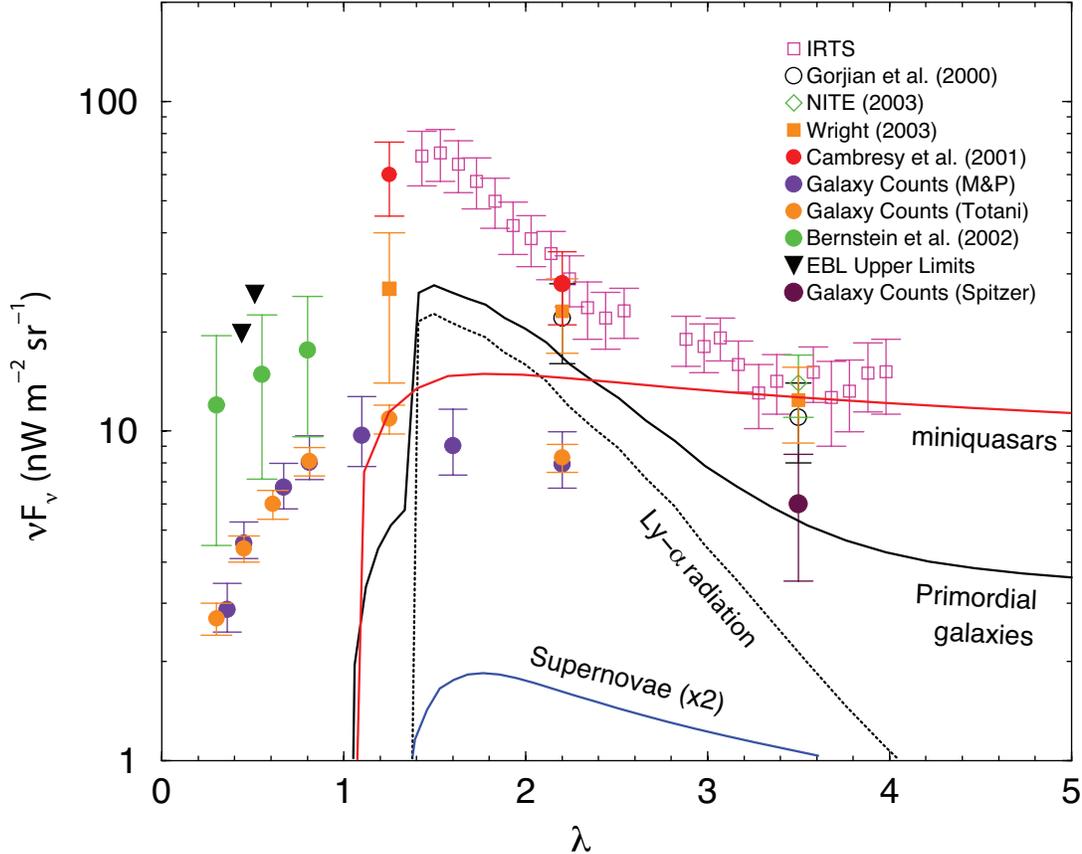,width=4.5in,angle=-90}}
\caption{Summary of observations of the extragalactic background at near-infrared
and optical wavelengths, showing upper limits, reported residuals after subtraction
of local foregrounds but without proof of isotropy (open symbols), reported detections,
and integrated galaxy counts.  The large experimental scatter at 1.2 $\mu m$ is
notable, the choice of zodiacal model being a prime culprit for systematic error.
The curves indicate model FLG-IRB components arising from ionizing radiation sources
at $z \sim 10-20$, including population III star formation in primordial galaxies,
mini-quasars, and supernovae}
\label{fig:irb}
\end{figure}

We compile recent measurements of the extragalactic near-infrared/optical background in
Figure~\ref{fig:irb} based on DIRBE \cite{Wright01}, \cite{Cambresy01}, \cite{Xu02}, \cite{Gorjian00}, \cite{Dwek98}, \cite{Hauser98}, IRTS \cite{Matsumoto05}, HST \cite{Bernstein02a}, and optical measurements \cite{Dube79}, \cite{Toller83} for
absolute photometery, and integrated galaxy number counts \cite{Totani01}, \cite{Madau00},
\cite{Fazio04}.

In the near-infrared and optical bands, the dominant foreground contaminant is zodiacal
light.  The IRTS and many DIRBE-based results use a model of IPD scattering and emission 
\cite{Kelsall98}.  Wright 2001 and Gorjian {\it et al.} (2000) use a different model based on
the principle that the zodiacal residual be zero at 25 $\mu m$ at high ecliptic latitude
\cite{Wright97}.  The optical measurement of Bernstein {\it et al.} 2002 subtracts zodiacal
emission based on the observed strength of scattered Fraunhofer lines from a ground-based
measurement \cite{Bernstein02b}.  Unfortunately, zodiacal dust models are not unique and
represent a leading source of systematic error.

Nevertheless, Figure~\ref{fig:irb} shows an intriguing trend:  the integrated galaxy
counts appear to fall well short of the IRB measured with absolute photometry in the near-infrared (particularly in the 1.2 and 2.2 $\mu m$ DIRBE bands), but are marginally consistent with the HST data at optical wavelengths, and with Spitzer at 3.6 $\mu m$.
Furthermore, the absolute photometry measurements appear to show a broad maximum at
1-2 $\mu m$.  This peak is either pronounced (the IRTS spectrum), or modest (the Wright
data points).  Clearly one must be duly cautious of systematic errors.  However, the
data may be hinting at the existence of a low-surface brightness component with a
spectrum peaked at 1-2 $\mu m$, which is the generic Ly-cutoff signature one expects
from first-light galaxies at high redshift.

\section{The Cosmic Infrared Background ExpeRiment}

DIRBE and the IRTS have carried out a comprehensive measurement of the infrared
sky brightness with absolute photometry, in the case of DIRBE involving highly
redundant all-sky measurements.  Nevertheless, key attributes of the IRB remain
unexplored, and can be addressed in a short sounding rocket experiment, needed to
escape strong airglow emission \cite{Ramsay92}, \cite{airglow}.  CIBER will search
for a first-light component to the IRB by studying the spatial anisotropy of
the IRB on angular scales from $7''$ to $2^\circ$.  A fluctuations experiment has
to potential to probe  for a first-light galaxy (FLG) IRB component to a level
roughly an order of magnitude fainter \cite{Cooray04a}, \cite{Kashlinsky04} than
has been realized with absolute photometry.  CIBER will study the absolute spectrum
of the IRB in the uncovered region between HST at 0.8 $\mu m$ and IRTS and
DIRBE at 1.2 $\mu m$.  Finally, CIBER will measure the intensity of the zodiacal
foreground using near-infrared Fraunhofer lines, allowing an independent check of
the Kelsall {\it et al.} zodiacal dust model.

\textbf{Wide-Field Imagers}

\begin{figure}[h]
\centerline{\psfig{file=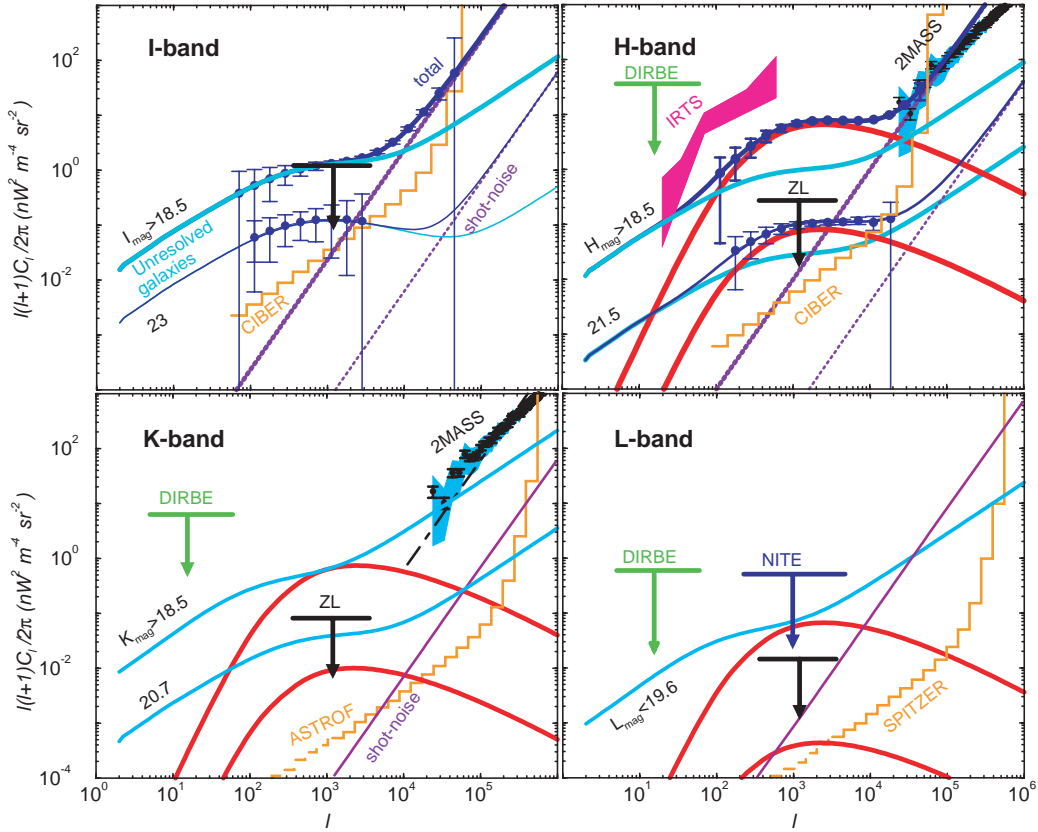,width=5.4in}}
\caption{Spatial power spectrum of the IRB in standard IR bands. The red
curves peaking at l = 1000 show the power spectra of first-light galaxy fluctuations from
$10 < z < 20$, with optimistic (top line) and pessimistic (bottom line) estimates. The
cyan curves give foreground galaxy clustering, the dotted lines show foreground galaxy
shot noise, as a function of magnitude cutoff.  The blue data points show the sensitivity of
CIBER on a single observation field, in multipole bins, including both statistical noise
and sample variance, on the optimistic and pessimistic models, using moderate and deep
source cutoffs.  The orange stair-steps show the statistical noise in log multipole
bins.  The results of previous experiments are indicated for DIRBE \cite{Kashlinsky00},
2MASS \cite{Kashlinsky02}, IRTS \cite{Matsumoto05}, and NITE \cite{Xu02}.  The orange
stair-steps, including foregrounds from their $3 \sigma$ point source cutoffs, show the
statistical sensitivities of Spitzer and ASTRO-F in their shallow survey fields, with dashed
lines showing where field mosaicing is needed.  Upper limits on zodiacal fluctuations
are shown, scaling from data with a $3^\prime$ beam at 25 $\mu m$ \cite{Abraham97} by the
zodiacal spectrum.}
\label{fig:psd}
\end{figure}

CIBER's imager optics are designed to detect FLG-IRB fluctuations from high redshift
$(z \sim 10-20)$ by producing deep, wide-field images on spatial scales from
$7''$ to $2^\circ$.  Fluctuations from foregrounds (e.g. zodiacal
light, galaxies) can be discriminated by using spatial spectra from the two bands
(see Figure~\ref{fig:psd}).  The imager data will be combined with shallow Spitzer and
ASTRO-F survey fields to allow us to detect and remove galaxies to a deeper level
than possible with CIBER alone.  Spitzer and ASTRO-F will make significant
contributions in this field (see \cite{Kashlinsky05}); however, CIBER images at
shorter wavelengths, which is especially important if the fluctuating component
shows a redshift-integrated Ly-cutoff signature.  Furthermore, CIBER is a wide-field
instrument optimized to detect FLG-IRB fluctuations on large angular scales, with
the wide multipole coverage needed to discriminate local-galaxy fluctuations.  By
observing the same fields as ASTRO-F and Spitzer, CIBER provides a powerful check
of systematic artifacts (e.g. flat fielding and mosaicing) between instruments, and
the combination gives comprehensive multi-color information.

The wide-field camera is a 10-cm refracting telescope cooled to 77 K, imaging
onto a $1024^{2}$ HgCdTe focal plane array and providing $7''$ pixels and a
$2^\circ \times 2^\circ$ field of view.  The detector array, sampled every 1.7 s
with a 25 s reset interval, provides background-limited performance.  A cold
shutter allows us to record dark frames prior to launch, and on descent after
observations have completed.  In the limited observation time available on a
sounding rocket flight, the cameras provide high sensitivity to diffuse emission,
with fidelity to signals on large angular scales.  Note that the sensitivity shown
in Table 1 and Figure~\ref{fig:psd} is for only a single 50 s field, and we
will have six such fields in a flight.

\begin{table}[h]
\centerline{
\begin{tabular}{|l|c|c|l|}
\multicolumn{4}{c}{\textbf{Table 1:  Imager Specifications}} \\
   \hline
Aperture & \multicolumn{2}{c|}{11} & cm \\
Pixel size & \multicolumn{2}{c|}{$7 \times 7$} & arcsec \\
FOV & \multicolumn{2}{c|}{$2 \times 2$} & degrees \\
	 \hline
$\lambda$ & 0.95(I) & 1.6(H) & $\mu m$ \\
$\Delta \lambda / \lambda$ & 0.5 & 0.5 & \\
Array QE & 0.65 & 0.75 & \\
Optics QE & 0.85 & 0.85 & \\
Photo current & 12 & 11 & $e^{-}/s$ \\
Dark current & $< 0.3$ & $< 0.3$ & $e^{-}/s$ \\
Read noise (CDS) & 15 & 15 & $e^{-}$ \\
$\delta \nu I_{\nu}$(inst) & 39 $(1 \sigma)$ & 19 $(1 \sigma)$ & $nW m^{-2} sr^{-1} pix^{-1}$ \\
		\hline
CIBER source cut & 18.6 ($3 \sigma$) & 18.0 ($3 \sigma$) & vega mag \\
 & 2.2e3 & 4.5e3 & sources $/$ sq degree \\
 		\hline
Deep galaxy cut & 23.0 & 21.4 & vega mag \\
 & 6e4 & 6e4 & sources $/$ sq degree \\
   \hline
\multicolumn{4}{l}{sensitivities quoted for a 50 s observation} \\
\end{tabular}}
\end{table}

CIBER reaches a limiting point source sensitivity of I = 18.5 and H = 17.8 mag
($3 \sigma$), somewhat deeper than the 2MASS all-sky survey.  Source confusion
\cite{Madau00} is 2-3 times fainter than the instrument noise level, assuming
that pixels are clipped at $3 \sigma$.  In order to remove the local-galaxy
clustering foreground more completely, deeper removal of point sources will be
necessary.  The pixel size of $7''$ is chosen to give us a sufficient density
of pixels to remove sources down to I = 23, H = 21.5 using ancillary ground-based
data.  CIBER pixels may be simply masked down to these cutoff magnitudes
with acceptable ($\sim 25\%$) data loss.

CIBER has the sensitivity to detect even the fluctuations from the pessimistic first-light galaxy model.  We note that the optimistic first-light
galaxy model assumes highly biased formation and contributes $\nu I_{\nu} = 25 nW m^{-2}
sr^{-1}$ to the IRB; the pessimistic model assumes moderately biased formation and
contributes $\nu I_{\nu} = 3 nW m^{-2} sr^{-1}$.  Thus a comprehensive
fluctuations measurement can probe for a first-light galaxy component to very low levels,
well below the current state of controversy in the IRB measurements (and see
\cite{Aharonian05}).

\textbf{Low-Resolution Spectrometer}

The low-resolution spectrometer is designed to measure the absolute spectrum
of the near-infrared background in the 0.8-2.0 $\mu m$ spectral region, in a
search for a redshifted Ly-cutoff feature from first-light galaxies.  As shown
in Figure~\ref{fig:irb}, the IRB may peak at 1-2 $\mu m$, and fall in brightness
in the optical.  The transition from the near-infrared to the optical, if
present, would have a much different spectral shape than either zodiacal
light or the integrated counts of galaxies, and could be detected without precise
removal of the zodiacal foreground.

The spectrometer is a 7.3 cm refracting telescope with re-imaging optics.  A prism
inserted in the collimated beam produces low-resolution spectra ($R \sim 20$, 
$\lambda = 0.8 - 2.0 \mu m$). We use a $256^{2}$ HgCdTe array with a pixel size of
40 $\mu m$, producing $1^{\prime} \times 1^{\prime}$ pixels and total field of view
of $4^{\circ}$.  Four slits located at the field stop produce $1^{\prime} \times
4^{\circ}$ strips on the sky, which the prism disperses into separate spectra on
the array.  As a result, 256 spectra are available for each slit, and a total of
1024 spectra are obtained.

\begin{table}[h]
\centerline{
\begin{tabular}{|l|c|l|}
\multicolumn{3}{c}{\textbf{Table 2:  Low-Resolution Spectrometer Specifications}} \\
   \hline
Aperture & 7.3 & cm \\
Pixel size & $1 \times 1$ & arcmin\\
FOV & $4 \times 4$ & degrees\\
Wavelength range & 0.8 - 2.0 & $\mu m$\\
Resolution & $\Delta / \Delta \lambda$ = 21-23 & \\
Slit size & $1 \times 256$ & arcmin\\
Optics QE & 0.8 & \\
Array QE & 0.5 & \\
Photo current & $10 \sim 20$ & $e^{-}/s$\\
Dark current & $< 0.1$ & $e^{-}/s$\\
Read noise (CDS) & $< 30$ & $e^{-}$\\
   \hline
\end{tabular}}
\end{table}

\begin{figure}[h]
\centerline{\psfig{file=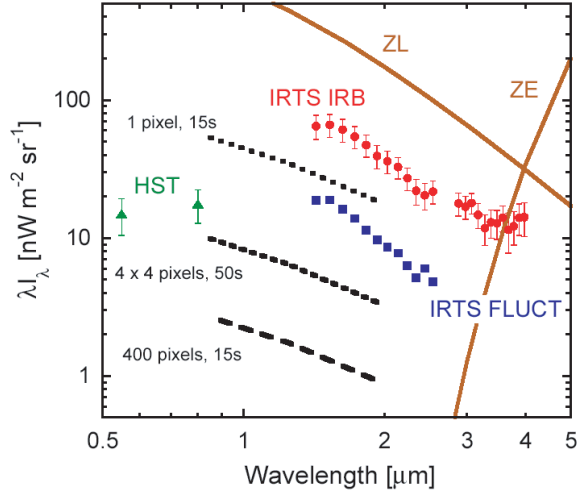,width=3.0in}}
\caption{Detection limits of the spectrometer at 1 $\sigma$ significance (dashed
lines) for a single pixel, co-addition of 4 spatial and 4 spectral pixels, and
co-addition of 400 spatial pixels (the entire field of view expected to be free
from point sources).  Shown in red and blue data points are the IRB and RMS
fluctuations reported by the IRTS.  The zodiacal light spectrum at the ecliptic
pole is shown for reference.}
\label{fig:lrs}
\end{figure}

The expected sensitivity of the spectrometer, which achieves background-limited
performance, is shown in Figure~\ref{fig:lrs} together with the zodiacal sky
brightness, and the IRB residual brightness and fluctuation spectra reported by
the IRTS.  The specifications of the spectrometer are summarized in Table 2.  In
order to obtain an accurate spectrum of the sky, we may co-add independent
spectra. After rejecting spectra that include point sources brighter than $18^{th}$
mag using the imager data, 400 such spectra can be co-added.  The resulting
sensitivity allows us to detect the absolute sky brightness with S/N $>$ 100
in only 15 s of integration time.  By co-adding four spatial and four spectral
pixels, we can detect IRTS/DIRBE fluctuations with $S/N \sim 5$, and we anticipate
having 300 such spatially independent pixels to this sensitivity.

\textbf{High-Resolution Spectrometer}

Uncertainties in zodiacal foreground removal play a leading role in the systematic
errors in determining the IRB, as evident by the large dispersion of data points
at 1.2 $\mu m$ in Figure~\ref{fig:irb}, which are the result of using different
zodiacal dust models based on DIRBE observations.

Prominent features in the solar spectrum can be used to trace the zodiacal light
since it is dominated by scattered sunlight at these wavelengths.  Narrow-band
spectroscopy, centered on and just off a Fraunhofer line, has been used
to estimate and remove the zodiacal light for optical extragalactic background
measurements \cite{Bernstein02a}, \cite{Dube79}, and to study the dynamics of
the zodiacal dust cloud \cite{Reynolds04}.  While simple in concept, this
technique is prone to its own set of systematic errors \cite{Mattila03}, most
notably absolute calibration, atmospheric extinction, atmospheric scattering
of zodiacal light and Galactic starlight, and even atmospheric scattering of
reflections off the ground.  Airglow emission greatly complicates measurements
at $\lambda > 0.55 \mu m$ from the ground.

With a small cooled spectrometer on a sounding rocket, we can measure near-IR
Fraunhofer features for direct comparison to the DIRBE zodiacal model, avoiding
the systematic errors associated with atmospheric scattering, and can absolutely
calibrate the instrument with a beam-filling source in the lab.

The narrow-band spectrometer is a 7.5 cm imaging refractor with a narrow-band
filter placed as the first optical element.  Rays passing through such a dielectric
etalon experience a wavelength shift $\lambda = \lambda_{0} \cos \theta_{i}$,
where $\theta_{i}$ is the angle of the ray inside the dielectric.  Thus the
wavelength response of the spectrometer varies over the array, by an amount
depending of the field of view.  This can be tailored so that the response at
the array center is tuned on the absorption feature, and shifts entirely off
the feature at the edge of the array.

\begin{table}[h]
\centerline{
\begin{tabular}{|l|c|c|l|}
\multicolumn{4}{c}{\textbf{Table 3:  High-Resolution Spectrometer Specifications}} \\
   \hline
Aperture & \multicolumn{2}{c|}{7.5} & cm \\
Pixel size & \multicolumn{2}{c|}{$2 \times 2$} & arcmin\\
FOV & \multicolumn{2}{c|}{$8.5 \times 8.5$} & degrees\\
Resolution & \multicolumn{2}{c|}{1000} & \\
Filter QE & \multicolumn{2}{c|}{0.7} & \\
Optics QE & \multicolumn{2}{c|}{0.9} & \\
Array QE & \multicolumn{2}{c|}{0.65} & \\
Photo current & \multicolumn{2}{c|}{1.5} & $e^{-}/s$\\
Dark current & \multicolumn{2}{c|}{$< 0.1$} & $e^{-}/s$\\
Read noise (CDS) & \multicolumn{2}{c|}{20} & $e^{-}$\\
		\hline
Wavelength & 0.8542 & 1.069 & $\mu m$\\
Line strength & 0.27 & 0.07 & nm\\
Contrast & 25 & 6 & $\%$ \\
$\nu I_{\nu}$(sky at NEP) & 550 & 450 & $nW m^{-2} sr^{-1}$ \\
$\delta \nu I_{\nu}$ & 55 $(1 \sigma)$ & 60 $(1 \sigma)$ & $nW m^{-2} sr^{-1} pix^{-1}$ \\
S$/$N zodi & 170 & 40 & $100 \times 100$ pix \\
$\Delta \nu I_{\nu}$ (zodiacal zero point) & 3 $(1 \sigma)$ & 12 $(1 \sigma)$ & $nW m^{-2}
sr^{-1}$ \\
   \hline
\multicolumn{4}{l}{sensitivities quoted for a 50 s observation} \\
\end{tabular}}
\end{table}

As shown in Table 3, the spectrometer can infer the zodiacal intensity with $S/N = 40$
observing a collection of lines at 1.07 $\mu m$, or $S/N = 170$ observing the 0.85 $\mu m$
CaII line, at the north ecliptic pole in a 50 s observation (and with higher S/N at
lower ecliptic latitudes).  The photon level, signal contrast, and the $S/N$ per
pixel are low, so careful attention must be paid to dark current, stray light,
and out-of-band blocking.  We will accurately measure the dark current in flight
using a cold shutter.

To compare with the Kelsall {\it et al.} zodiacal dust model, we must use the measured
zodiacal color from the low-resolution spectrometer to extrapolate the zodiacal signal
to the 1.2 and 2.2 $\mu m$ DIRBE bands.  The low-resolution spectrometer data are
essential because the zodiacal spectrum departs from the solar spectrum around $1 \mu m$,
and has not been accurately measured at $\lambda < 1.2 \mu m$.

Finally we note the Fraunhofer line measurements provide an absolute tracer of the
zodiacal foreground, and combined with the low-resolution spectrometer, allow a
determination of the absolute infrared background from 0.8 - 2.0 $\mu m$.

\section{Summary}

Because CIBER has multiple capabilities, we summarize the measurement objectives.
CIBER will:

\begin{itemize}
\item Probe the IRTS fluctuations in several fields with $100 \times$ higher 
signal-to-noise, using spatial and color information to determine if the IRTS 
fluctuations arise from zodiacal, Galactic, local-galaxy clustering, or first-light
galaxy origins.

\item Use CIBER's $100 \times$ higher sensitivity, $100 \times$ higher angular
resolution, and $500 \times$ deeper point source removal than IRTS to probe the
spectrum of IRB fluctuations for a FLG-IRB component at a level $100 \times$ fainter
than the IRTS fluctuations.  Combine CIBER data with ASTRO-F and Spitzer surveys
for the definitive data set for IRB fluctuations in 4 bands from 0.9 to 3.6 $\mu m$.

\item Search for the Lyman cutoff signature of reionization in the IRB between
the optical and near-infrared IRB measurements.

\item Measure the zodiacal foreground in several fields using a narrow-band
spectrometer to trace scattered near-infrared Fraunhofer absorption features.
Current zodiacal model discrepancies based on DIRBE are at the $\sim 10 \%$
level, producing large discrepancies in the derived IRB.

\item Combine the low-resolution and narrow-band spectrometer data to probe the IRB
from 0.8 - 2.0 $\mu m$ down to the narrow-band spectrometer's zodiacal subtraction
limit.

\end{itemize}

The authors wish to acknowledge the support of NASA grant NNG05WC18G.

\end{document}